\documentclass[fleqn,twoside]{article}
\usepackage{espcrc2}
\usepackage{subfigure}
\usepackage{amsmath}
\usepackage{lineno}
\usepackage{graphicx}
\usepackage[figuresright]{rotating}
\begin{document}

\title{Measurements of cosmic-ray energy spectra 
 with the 2$^{nd}$ CREAM flight
}

\author{
P. Maestro\address[4]{\footnotesize{Department of Physics, University of Siena and INFN, Via Roma 56, 53100 Siena, Italy}}\thanks{Corresponding author. \newline e-mail address:\emph{paolo.maestro@pi.infn.it} (P. Maestro).},
H. S. Ahn\address[1]{\footnotesize{Institute for Physical Science and Technology, University of Maryland, College Park, MD 20742, USA}}, 
P. Allison\address[3]{\footnotesize{Department of Physics, Ohio State University, Columbus, OH 43210, USA}}, 
M. G. Bagliesi\addressmark[4], 
L. Barbier\address[10]{\footnotesize{Astroparticle Physics Laboratory, NASA Goddard Space Flight Center, Greenbelt, MD 20771, USA}}, 
J. J. Beatty\addressmark[3], 
G. Bigongiari\addressmark[4], \\
T. J. Brandt\addressmark[3], 
J. T. Childers\address[5]{\footnotesize{School of Physics and Astronomy, University of Minnesota, Minneapolis, MN 55455, USA}}, 
N. B. Conklin\address[6]{\footnotesize{Department of Physics, Penn State University, University Park, PA 16802, USA}},
S. Coutu\addressmark[6], 
M. A. DuVernois\addressmark[5],
O. Ganel\addressmark[1], 
J. H. Han\addressmark[1], \\
J. A. Jeon\address[7]{\footnotesize{Department of Physics, Ewha Womans University, Seoul 120-750, Republic of Korea}}, 
K. C. Kim\addressmark[1],
M. H. Lee\addressmark[1], 
A. Malinine\addressmark[1], 
P. S. Marrocchesi\addressmark[4], 
S. Minnick\address[8]{\footnotesize{Department of Physics, Kent State University, Tuscarawas, New Philadelphia, OH 44663, USA}}, 
S. I. Mognet\addressmark[6], \\
S. W. Nam\addressmark[7], 
S. Nutter\address[9]{\footnotesize{Department of Physics and Geology, Northern Kentucky University, Highland Heights, KY 41099, USA}} ,
I. H. Park\addressmark[7], 
N. H. Park\addressmark[7], 
E. S. Seo\addressmark[1]\address[2]{\footnotesize{Department of Physics, University of Maryland, College Park, MD 20742, USA}}, 
R. Sina\addressmark[1], 
P. Walpole\addressmark[1], 
J. Wu\addressmark[1], \\
J. Yang\addressmark[7], 
Y. S. Yoon\addressmark[1]\addressmark[2], 
R. Zei\addressmark[4],  
S. Y. Zinn\addressmark[1]
}
\begin{abstract}
 During its second Antarctic flight, 
the  CREAM  (Cosmic Ray Energetics And Mass) balloon experiment
collected
 data for 28 days, measuring the charge and the energy of cosmic rays (CR)  
 with a redundant system of particle identification
 and an imaging thin ionization calorimeter.
Preliminary direct measurements of the absolute intensities 
of individual CR nuclei 
are reported in the elemental range from carbon to iron at very high energy.\\\\
\vspace{-0.6cm}
\end{abstract}
\maketitle
\section{Introduction}
The CREAM experiment was designed to measure 
the composition and  energy spectra of 
cosmic rays 
approaching energies up to 10$^{15}$ eV. 
Since December 2004, four instruments were successfully flown on balloons 
over Antarctica 
where they collected
several  million CR events in the elemental range from hydrogen to iron,
 and with total particle energies
reaching the 100 TeV scale and above. The final goal of the experiment
is to provide a deeper understanding of the acceleration mechanism  of cosmic rays
and to test the validity of the astrophysical models describing their propagation  in the Galaxy \cite{1}.\\
In this paper, we present  preliminary energy spectra of the  
even-charged, abundant nuclei from carbon to iron as measured by the instrument during its second flight (CREAM-II).
\section{The CREAM-II instrument}
The instrument for the second flight included: a
redundant system for particle identification, consisting 
(from top to bottom)
of a timing-charge detector (TCD), a
Cherenkov detector (CD), a pixelated silicon charge detector (SCD), and  
a sampling imaging calorimeter (CAL) designed to provide a
measurement of the energy of primary nuclei in the multi-TeV region. \\
The TCD is comprised of two planes (120 $\times$ 120 cm$^2$) 
of four 5 mm-thick plastic scintillator paddles, 
read out 
by fast timing photomultiplier tubes (PMT). 
The CD is a 1 cm-thick plastic radiator, with 1 m$^2$ surface area, read out by eight PMTs via wavelength shifting bars. 
The SCD is a dual layer made of 312 silicon sensors, each segmented as an array of 4$\times$4 
pixels, which
covers an effective area of 0.52 m$^2$ with no dead regions.\\
The CAL is a stack of 20 tungsten plates (50$\times$50  cm$^{2}$, 
each 1 X$_0$ thick) with interleaved active layers 
of
1 cm-wide scintillating fiber ribbons, read out by 40 hybrid photodiodes (HPD).
It is preceded by 
a 0.47 $\lambda_{int}$-thick graphite target
to induce hadronic interactions of CR nuclei.\\
The second CREAM payload was launched on December 16$^{th}$  2005 from McMurdo  and
flew over Antarctica until January 13$^{th}$  2006,  at a
balloon float altitude  between 35 and 40 km.
A detailed description of the instrument and its flight performance can be found in \cite{Marrocchesi}.
\section{Data analysis}
A data set collected in the period Dec. 19$^{th}$- Jan. 12$^{th}$, under stable instrument conditions,  
was used in this analysis.
The first step of the procedure consists of selecting events with an accurate 
trajectory reconstruction of the primary particle, and measuring its charge and energy.
The direction of the particle is given by the axis of the shower 
reconstructed in the imaging calorimeter. This is obtained by a $\chi^2$ fit of the candidate
track points sampled along the longitudinal development of the shower.
On each CAL plane, they are
defined as the center of gravity of the cluster
formed by the cell with maximum signal and its two neighbours.  
The fitted shower axis is back-projected to the target and its intersections
with the two SCD layers define the impact points
of the primary particle, with a spatial resolution better than 1 cm rms.
In each SCD plane, the pixel with the highest signal is located 
inside a circle of confusion with a 3 cm radius 
centered on the impact point. \\
\begin{figure}
\begin{center}
\includegraphics[scale=0.38]{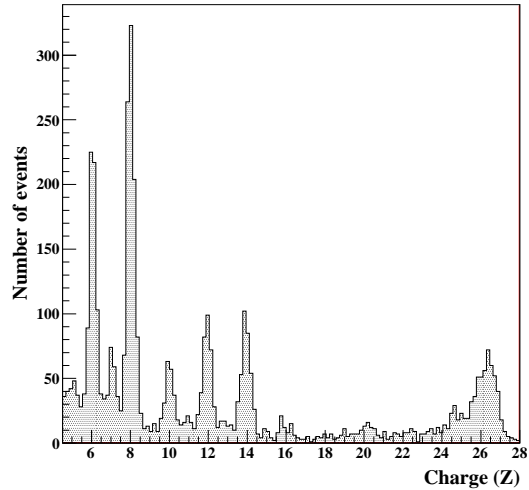}
\caption{Charge histogram obtained by the SCD in the elemental range from boron to iron.}
\label{fig1}                            
\end{center}
\vspace{-0.9cm}
\end{figure}
If the signals of the matched hits in the upper and lower 
SCD planes are consistent within 30\%, 
they are selected as 
two independent samples of the specific ionization $dE/dx$ 
and
corrected for the pathlength 
traversed by the particle in the silicon sensors. 
A charge $Z$ is assigned to the particle by averaging the two measured values
of the specific ionization 
and taking into account its $Z^2$ dependence.
The charge distribution reconstructed by the SCD is shown in figure \ref{fig1};
by fitting each peak to a gaussian, a charge resolution 
$\sigma$ is estimated (in units of the electron charge $e$) as:
0.2 for C, N, O; $\sim$ 0.23 for Ne, Mg, Si; $\sim$ 0.5 for Fe.
A 2$\sigma$ cut around the mean charge value is applied to select samples of each element, while for iron a
1$\sigma$ cut is used. In this way, 1288 O are retained, as well as 456 Si 
and 409 Fe candidates. 

The total energy ($E_d$), deposited in the calorimeter by an interacting nucleus,
is measured by summing up the calibrated signals of its cells. 
In order to infer the primary particle energy $E$ from $E_d$, an unfolding procedure is applied.
In fact, due to the finite energy resolution of the detector, the measured counts  
in a given energy interval must be corrected for overlap with the neighbouring bins. 
\begin{figure}
\begin{center}
\includegraphics[scale=0.4]{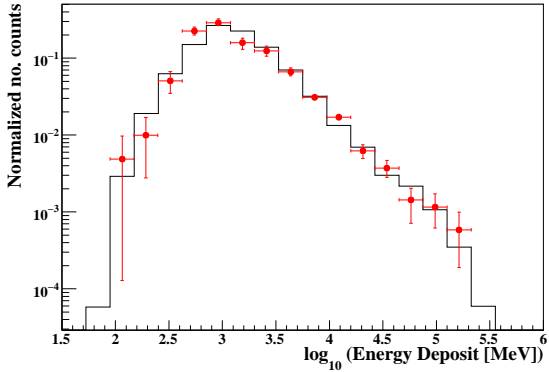}
\caption{Energy deposited in the calorimeter by a selected sample of carbon nuclei. Simulated (histogram) and real (dots)
events are shown.}
\label{fig2}                            
\end{center}
\vspace{-0.7cm}
\end{figure}
This can be done by inverting the matrix equation: 
$M_i = \sum_j A_{ij} N_j$ \,\,
where $N_j$ and $M_i$ are  the ``true'' and measured
counts in each energy bin, respectively. A generic element of the mixing matrix $A_{ij}$ represents 
the probability that a CR particle, carrying an energy 
corresponding to a given energy  bin \emph{j}, produces 
an energy deposit in the calorimeter falling in the bin \emph{i}.
A detailed MonteCarlo (MC) simulation of the instrument,
based on the FLUKA 2006.3b package 
\cite{fluka}, was developed
 to estimate the unfolding matrix. 
Sets of nuclei, generated isotropically and with energies chosen according to a power-law spectrum,
are analyzed with the same procedure as that 
used for the flight data. Each matrix element $A_{ij}$
is calculated by correlating the generated spectrum with the distribution of the 
deposited energy in the calorimeter \cite{Zei}.
In order to get a reliable set of values of the unfolding  matrix for each nucleus, 
the MC simulation is finely tuned to reproduce both flight data and the calibration data
collected with accelerated particle beams \cite{Ahn}.
The agreement of the MC description with the real instrument behaviour was carefully checked.
As an example, in figure \ref{fig2}
the response of the calorimeter to carbon nuclei from the flight data is compared
with an equivalent set of simulated events.
\section{Energy spectrum}
The interval of energy spanned by the measured CR events
is divided into bins of width $\Delta E_i$, larger than 
the energy resolution
of the calorimeter 
and centered at values $E^{med}_i$ (calculated according to the definition of \cite{LW}).
For each element,  the ``true'' number of counts $N_i$
in each bin is obtained as a result of the unfolding algorithm. 
The absolute differential intensity $\Phi$ at an energy $E^{med}_i$ is calculated according to the formula 
\begin{equation*}
\Phi(E^{med}_i) = \frac{N_i}{\Delta E_i}\times \frac{1}{\epsilon \times \text{TOI} \times \text{TOA} \times S\Omega \times \text{T} }
\end{equation*}
where 
T is the exposure time, $\epsilon$ the efficiency of the selection cuts, $S\Omega$ the geometric factor
of the instrument, and TOI and TOA are, respectively, the corrections to the top of instrument
and to the top of the atmosphere.\\
The geometric factor $S\Omega$ is estimated from MC simulations
 by counting the fraction of generated particles 
entering the upper SCD plane and crossing the upper CAL plane. A value of 0.46 m$^2$sr is found. 
During the flight, the live time T
was measured by 
the housekeeping system onboard. 
The selected set of data amounts to a live time of 
16 days and 19 hours, close to 75\% of the real time of data taking.\\
The overall efficiency $\epsilon$ is estimated from MC simulations; 
it has a constant value of around 70\% at energies $>$
3 TeV for all nuclei.\\
The probability that a nucleus undergoes a spallation reaction in the 
amount of material  ($\sim$ 4.8 g/cm$^2$) above
the upper SCD plane is also estimated from MC simulations.
The fraction of surviving nuclei, i.e., the TOI correction, 
spans from 81.3\% for C 
to 61.9\% for Fe.
The TOA correction is calculated  by simulating with FLUKA  the atmospheric
 overburden
during the flight (3.9 g/cm$^2$ on average).
Survival probabilities
ranging from 84.2\% for C to 
71.6\%  for Fe  are found.
\section{Results}
\begin{figure*}[t]
\begin{center}
\includegraphics[scale=0.78]{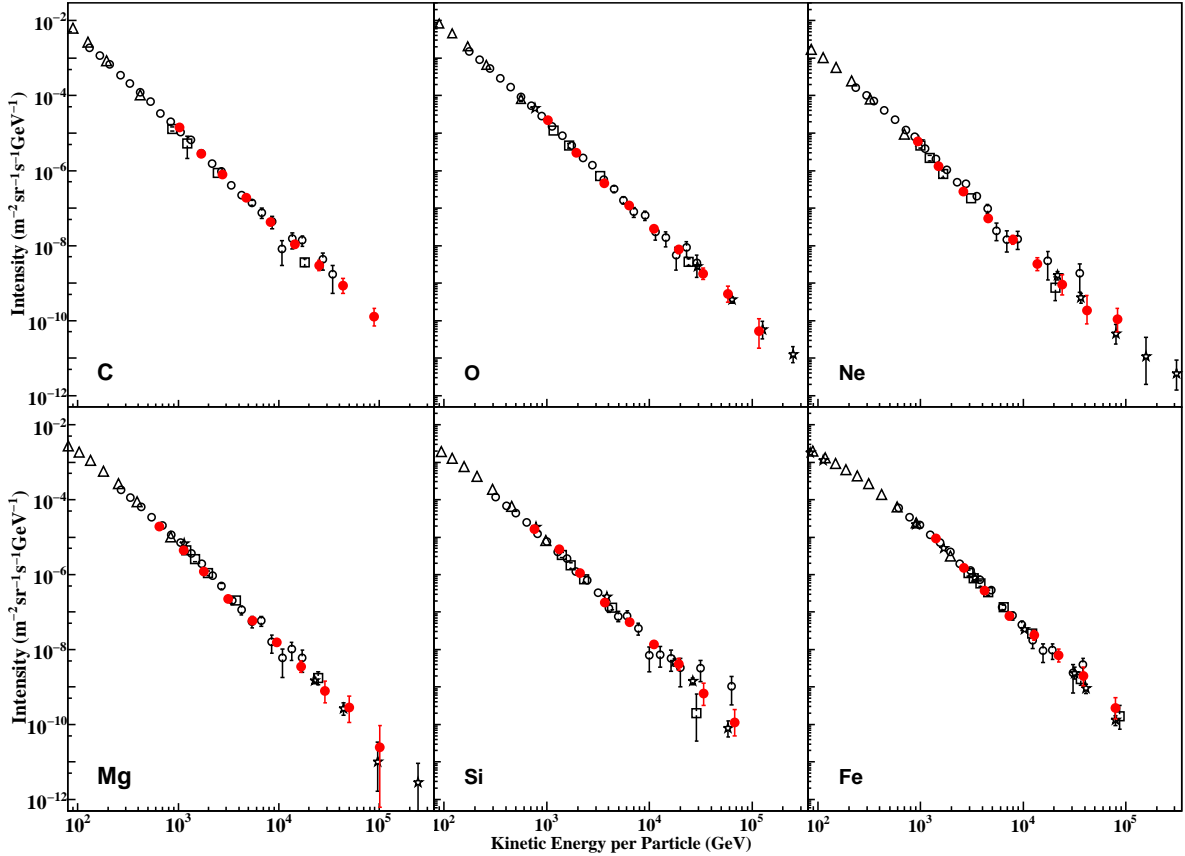}
\caption{Energy spectra 
of the more abundant heavy nuclei. 
Results of CREAM-II (filled circles)
are compared with measurements from HEAO \cite{HEAO} (triangles), CRN \cite{CRN} (squares), ATIC \cite{ATIC} (open circles) 
and TRACER \cite{TRACER} (stars).
}
\label{fig3}                            
\end{center}
\vspace{-0.5cm}
\end{figure*}
The preliminary energy spectra of C, O, Ne,  Mg, Si and Fe  
measured by CREAM-II are shown in 
figure \ref{fig3}. Only statistical errors are reported, a detailed
study of the systematic uncertainties being underway.
Absolute particle intensities are presented without any arbitrary normalization to previous data.
The particle energy range extends from around 800 GeV up to 100 TeV. 
CREAM-II data are found  in general to be in good agreement with measurements of
previous balloon-borne and satellite experiments \cite{HEAO,CRN,ATIC,TRACER}.
Though still preliminary, they seem to suggest that
the intensities of the more abundant heavy elements 
have a very similar energy dependence.
A more refined analysis, including an assessment of the systematics, is still in progress. 
\section{Conclusion}
The CREAM-II instrument carried out measurements of high-Z  cosmic-ray nuclei
with an excellent charge resolution and 
a reliable determination of their energy.
Energy spectra of the more abundant heavy nuclei
are measured and found to agree well with other direct measurements. 

\end{document}